\documentclass{article}
\usepackage{spconf,amsmath,graphicx,hyperref}
\usepackage{amsfonts}
\usepackage[capitalize]{cleveref}
\usepackage{booktabs}
\usepackage{makecell}
\usepackage{multirow}
\usepackage{tabularx}
\usepackage{graphicx}
\usepackage[activate]{microtype}
\usepackage[acronym, shortcuts, nohypertypes={acronym}]{glossaries}
\newacronym{CLAP}{CLAP}{contrastive language-audio pretraining}
\newacronym{CASA}{CASA}{computational audiotiry scene analysis}
\newacronym{CP}{CP}{computational paralinguistics}
\newacronym{SER}{SER}{speech emotion recognition}
\newcommand{\sclap}{SmoothCLAP }

\title{SmoothCLAP: Soft-Target Enhanced Contrastive Language\--Audio Pretraining for Affective Computing}
\name{\em Xin Jing$^{\star}$ \qquad
      Jiadong Wang$^{\star}$ \qquad
      Andreas Triantafyllopoulos$^{\star}$ \qquad
      Maurice Gerczuk$^{\star}$ \qquad\\\em Shahin Amiriparian$^{\dagger\star}$ \qquad
      Jun Luo$^{\dagger}$ \qquad
      Björn Schuller$^{\star\ddagger}$}

\address{$^{\star}$ CHI -- Chair of Health Informatics, TUM University Hospital, Munich, Germany, \\
         $^{\dagger}$ Huawei, Netherlands, $^{\ddagger}$GLAM, Imperial College London, UK}

\begin{document}
\ninept
\maketitle
\begin{abstract}
The ambiguity of human emotions poses several challenges for machine learning models, as they often overlap and lack clear delineating boundaries.
Contrastive language-audio pretraining (CLAP) has emerged as a key technique for generalisable emotion recognition.
However, as conventional CLAP enforces a strict one-to-one alignment between paired audio–text samples, it overlooks intra-modal similarity and treats all non-matching pairs as equally negative. 
This conflicts with the fuzzy boundaries between different emotions.
To address this limitation, we propose SmoothCLAP, which introduces softened targets derived from intra-modal similarity and paralinguistic features. 
By combining these softened targets with conventional contrastive supervision, SmoothCLAP learns embeddings that respect graded emotional relationships, while retaining the same inference pipeline as CLAP.
Experiments on eight affective computing tasks across English and German demonstrate that SmoothCLAP 
is 
consistently achieving superior performance. Our results highlight that leveraging soft supervision is a promising strategy for building emotion-aware audio–text models.
\end{abstract}
\begin{keywords}
computational paralinguistic, speech emotion recognition, contrastive learning, zero-shot learning
\end{keywords}

\section{Introduction}
\label{sec:intro}
As a subfield of affective computing, \ac{CP} analyses vocal phenomena such as prosody, intonation, and spectral characteristics, and their relation to speaker traits and states~\cite{Schuller14-CPE}. These cues have been applied not only to the recognition of ``classic'' emotions, but also to a broad range of speaker-related dimensions, including personality, likability, sincerity, deception, and health indicators~\cite{Zhang24-PAL}. Beyond their role in affect recognition, CP features constitute intrinsic audio attributes that can serve as auxiliary targets to enrich and diversify affective speech corpora~\cite{Jing25-EET}.
However, one persistent challenge is the need for task-specific labeled data~\cite{Wagner23-DOT}.
Recently, the \ac{CLAP} framework~\cite{Elizalde23-CLA} has emerged as a general-purpose alternative, establishing a new paradigm for learning audio representations by aligning them with textual descriptions in a shared latent space~\cite{Triantafyllopoulos25-CAF}. Unlike conventional supervised, label-dependent methods, CLAP enables flexible, zero-shot learning guided by natural language, and has shown strong potential for affective computing by modeling emotional audio–text relationships beyond predefined categorical labels.
Building on CLAP, ParaCLAP~\cite{Jing24-PTA} introduced a novel and systematic process for generating descriptive text queries, effectively translating expert acoustic knowledge into the natural language domain. While ParaCLAP successfully addressed the data generation problem, EmotionRankCLAP~\cite{Chandra25-EBN} extended CLAP by embedding the ordinal nature of emotions -- modeled via continuous valence-arousal dimensions -- into the contrastive learning objective, encouraging embeddings to reflect emotion-relative distances rather than binary correspondence. 
While these works illustrate the promise of CLAP by addressing descriptive query generation and intra-emotion variability, its learning process still relies on bringing paired audio–text samples closer while treating all non-corresponding pairs as equally negative. 
This assumption overlooks the underlying structure of affective data, where emotional annotations and data often exhibit fuzzy boundaries and many-to-many relationships~\cite{Triantafyllopoulos24-BDL}. For instance, `disgust' and `fear' are conceptually closer than `happiness' and `fear'.
Consequently, enforcing a strict separation between all non-corresponding pairs creates a gap between the training target and the underlying structure of the problem, where the geometric structure of the embedding space does not accurately reflect the psychological structure of emotion~\cite{Chandra25-EBN, Gao24-SSC}.

To address this issue, we propose \sclap, a framework that integrates computational paralinguistics into CLAP training via soft-target supervision. 
Specifically, starting from classic paralinguistic features and the set of labels available for a large \ac{SER} dataset, we construct a diverse set of (audio, tags) samples. Tags are leveraged as auxiliary signals to construct softened targets that replace rigid one-hot labels, allowing the model to capture more flexible relationships among samples. This design enriches the training signal with instructive information from CP features while avoiding the instability of purely similarity-based targets. Importantly, \sclap introduces these modifications only at the training stage; the inference procedure remains identical to CLAP, ensuring compatibility and efficiency. Experimental results demonstrate that incorporating CP-driven softened targets improves recognition performance in affective computing tasks.
Our main contributions are summarized as follows:

\begin{itemize}
    \item We introduce intra-modal self-similarity from CP features as softened supervision signals, addressing the strict one-to-one assumption in standard CLAP.
    \item By exploiting CP features as auxiliary tags, \sclap enriches the embedding space with paralinguistic priors, enhancing the diversity and semantic richness of supervision.
    \item we introduces a novel CLAP object by fusing the intra-modal similarity with the conventional one-hot supervision, improving alignment between affective multimodal embeddings.
\end{itemize}

The remainder of this paper is organized as follows.
we describe the proposed \sclap\ framework, including the integration of CP features and the design of soft-target supervision in \cref{sec:met}.
In \cref{sec:exp}, we present the experimental setup, ablation studies, and results, followed by a detailed analysis.
Finally, in \cref{sec:con}, we summarize our findings and discuss future directions.

\begin{figure*}
    \centering
    \includegraphics[width=\linewidth]{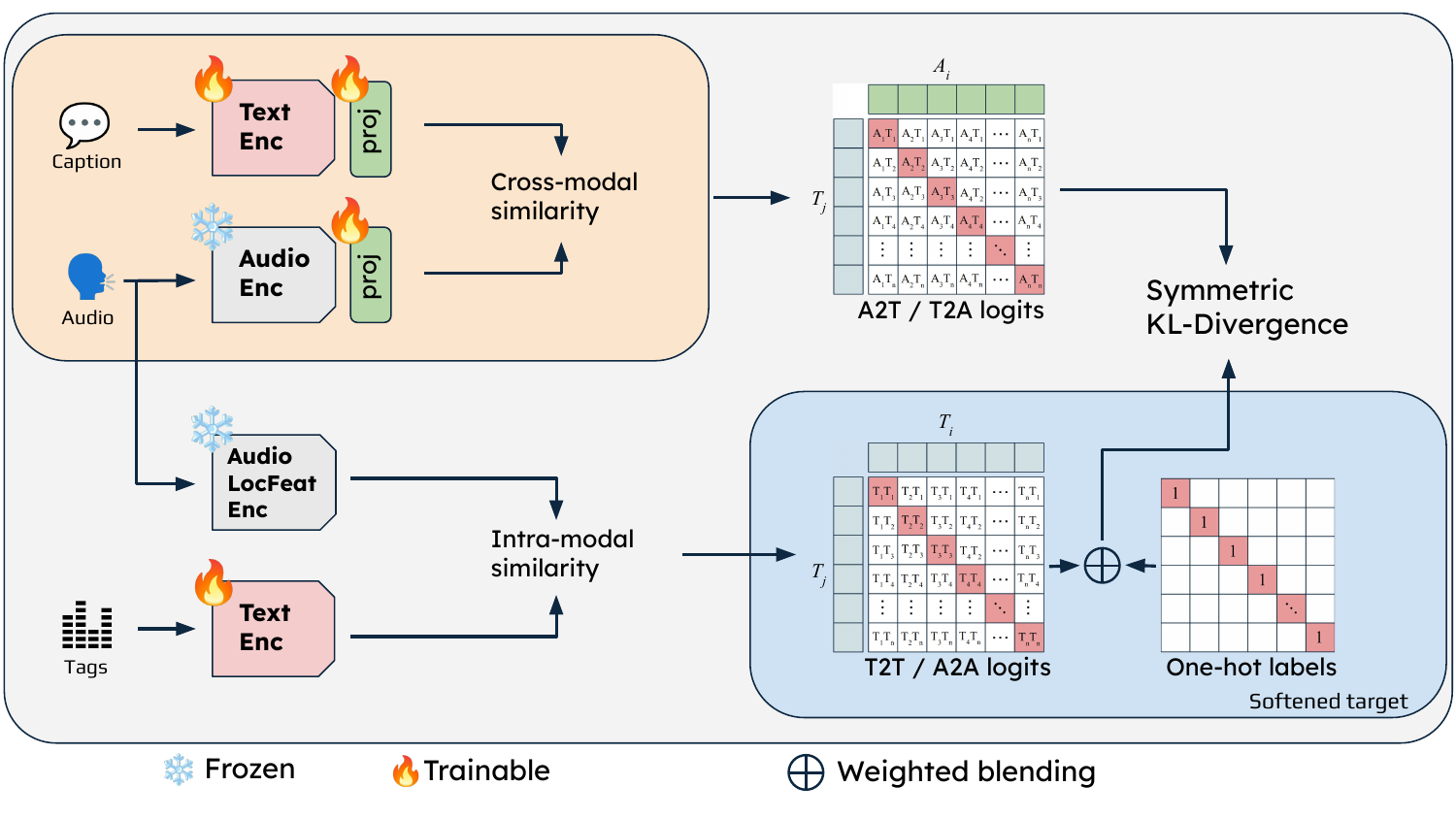}
    \caption{Architecture of SmoothCLAP. The model integrates intra-modal self-similarity to enable soft-target smoothing. It promotes alignment across modalities while preserving semantic relationships within both the audio and text domains.}
    \label{fig:framework}
\end{figure*}

\section{Methodology}
\label{sec:met}
\textbf{CLAP Preliminaries: }
Let $f_a(\cdot)$ and $f_t(\cdot)$ denote the audio and text encoders. Consider audio-text pairs ${(A_i, T_i)}$, the embedding of encoders are defined as:

\begin{equation}
    z_i^a=proj_A(f_a(A_i)), \quad z_i^t=proj_T(f_t(T_i)),
\end{equation}

$proj_A$ and $proj_T$ denote as the projection layer for audio and text respectively that will map the both embeddings to one shared latent space. Then, CLAP measures the similarity by a scaled cosine similarity:

\begin{equation}
    Sim=\tau * (z_i^a\cdot z_i^t),
\end{equation}
where $\tau$ is a learnable temperature parameter.
The CLAP object is to encourage audio-text pairs to be similar while push unpaired embeddings apart. The symmetric InfoNCE loss can be expressed as:
\begin{equation}
    L_{CLAP} = 1/2 * (H_t(S) + H_a(S)),
\end{equation}
where $h_k = \frac{1}{N}\sum_{n=0}^Nlog(diag(softmax(Sim))$ along the audio and text axis.

\noindent
\textbf{SmoothCLAP: }
According to the \cref{fig:framework}, the overall architecture of \sclap follows the conventional CLAP design, which also consists of an audio encoder and a text encoder. Each encoder output is projected into a shared embedding space via learnable projection layers. However, during the training phase, negative samples in CLAP's loss are treated with uniform one-hot supervision. This rigid target distribution ignores the inherent structure within each modality, and can hinder generalization in domains where samples exhibit graded similarities. To address this, we introduce \textit{intra-modal self-similarity} to retain soft-target smoothing and encourage the model to align with cross-modal correspondences while respecting semantic relationships within the audio and text modalities.

For the audio branch, \sclap obtains the local frame features extracted by a pretrained local feature extractor. By computing pairwise similarities between the aggregated local features across samples within a batch, we obtain an audio-to-audio (a2a) similarity matrix, which encodes the relational structure of the audio domain itself. This intra-modal similarity distribution is then employed as a component of the soft target, providing additional supervision that reflects the inherent structure of the audio space. The similar processing is applied to caption and auxiliary tags, and the only difference for the text branch is that we apply the same text encoder to extract the text embeddings and then a text-to-text (t2t) similarity matrix. 
By introducing softened targets, \sclap transfers a small portion of confidence to negatives, thereby also allowing for a degree of similarity within non-matching pairs. While such smoothing captures a naive many-to-many relationship, it does not leverage the rich relational structures inherently present within a single modality.

In this work, we employ a pretrained wav2vec 2.0 large model fine-tuned for dimensional \ac{SER}\footnote{https://huggingface.co/audeering/wav2vec2-large-robust-12-ft-emotion-msp-dim}~\cite{Wagner23-DOT}
 as both the audio encoder and the local feature extractor. The pretrained model was pruned from 24 to 12 transformer layers prior to fine-tuning on the MSP-Podcast (v1.7) dataset~\cite{lotfian2017building}.
 We also experiment with additional feature extractors as part of our ablation study.
For the text encoder, we use the BERT base model (uncased)~\cite{Devlin18-BPO}; the [CLS] token from the final layer is used as the text embedding.
Given a minibatch of $B$ audio--text pairs $\{(a_i, t_i)\}_{i=1}^B$, 
we define the cross-modal similarity score as:
\begin{equation}
    s_{ij} = \frac{e^a_i \cdot e^t_j}{\tau},
\end{equation}
where $e^a_i, e^t_j \in \mathbb{R}^d$ are the $\ell_2$-normalized audio and text embeddings, 
and $\tau$ is a learnable temperature parameter.

Instead of using one\--hot labels, SmoothCLAP constructs soft distributions from intra-modal similarities

\begin{align}
    q^{a2a}_{ij} &= \frac{\exp(\bar{\ell}^a_i \cdot \bar{\ell}^a_j / \tau_{a2a})}
    {\sum_{k=1}^B \exp(\bar{\ell}^a_i \cdot \bar{\ell}^a_k / \tau_{a2a})}, \\
    q^{t2t}_{ij} &= \frac{\exp(e^t_i \cdot e^t_j / \tau_{t2t})}
    {\sum_{k=1}^B \exp(e^t_i \cdot e^t_k / \tau_{t2t})},
\end{align}

where $\bar{\ell}^a_i$ is the mean-pooled local audio representation of sample $i$.  
The combined soft target distribution is:
\begin{equation}
    q_{ij} = (1-\gamma) q^{a2a}_{ij} + \gamma q^{t2t}_{ij},
\end{equation}
with `mix gamma' $\gamma \in [0,1]$ controlling the weighting between audio- and text-side similarities.
The final soft target distribution is a mixture of a2a and t2t similarities:

\begin{equation}
    y_{ij} = (1-\beta)\delta_{ij} + \beta q_{ij},
\end{equation}
where $\delta_{ij}$ is the identity matrix, $\beta$ is the fusion factor to control the trade-off between hard and softened targets.
The model’s predicted distributions are defined as:
\begin{align}
    p^{a2t}_{ij} &= \frac{\exp(s_{ij}/\tau_{\text{pred}})}
    {\sum_{k=1}^B \exp(s_{ik}/\tau_{\text{pred}})}, \\
    p^{t2a}_{ij} &= \frac{\exp(s_{ji}/\tau_{\text{pred}})}
    {\sum_{k=1}^B \exp(s_{jk}/\tau_{\text{pred}})}.
\end{align}

The soft-target loss is formulated as a symmetric KL divergence between 
the predicted distributions and the softened targets:
\begin{equation}
    \begin{split}
        \mathcal{L}_{\text{soft}} = \frac{1}{2B} \sum_{i=1}^B \Big[ 
        & \, KL(y_i \;\|\; p^{a2t}_i) + KL(p^{a2t}_i \;\|\; y_i) \\
        & + KL(y_i \;\|\; p^{t2a}_i) + KL(p^{t2a}_i \;\|\; y_i) 
        \Big].
    \end{split}
\end{equation}

Notably, while \sclap leverages intra-modal similarity as auxiliary supervision during training, its inference phase is identical to that of CLAP.

\noindent
\textbf{Tag Generation: }
For generating tags, we followed the same protocol as our previous work~\cite{Jing24-PTA}.
The tags for each speech sample are derived from two main sources:
1) dataset-provided labels, and
2) CP features extracted from the audio sample.
In both cases, dimensional attributes are discretized through binning and subsequently converted into textual descriptors.

\textbf{Dataset labels:}
The training dataset contains categorical emotion labels, dimensional emotion ratings (arousal, valence, dominance), and gender information. For the dimensional attributes, values are binned according to their empirical distribution (bottom 30\,\%, middle 40\,\%, top 30\,\%) and mapped to template phrases (e.g.,``low/mid/high arousal'').

\textbf{Tags from expert features:}
We further extract acoustic features using eGeMAPS~\cite{Eyben15-TGM, Eyben10-OTM}. Although eGeMAPS includes 88 parameters, we focus on a subset that is more interpretable including mean ($\mu$) and standard deviation ($\sigma$) of \emph{pitch}, \emph{intensity}, \emph{jitter}, \emph{shimmer}, and total utterance \emph{duration}. Each feature is binned using the same distribution-based thresholds as the dimensional variables, and tags are generated accordingly (e.g., ``low/normal/high pitch'').

\section{Experiments \& Results}
\label{sec:exp}
\subsection{Experimental Setup}
\textbf{Training dataset:} Our training dataset is MSP-Podcast v1.9~\cite{lotfian2017building}, containing over 110 hours of English speech extracted from podcasts. The corpus includes 55\,283 utterances produced by more than 1\,200 speakers and is annotated via crowd-sourcing with 10 emotion categories. For training, we adopted the commonly used category selection, in which utterances labeled as `no\_agreement' were excluded, resulting in a final set of 45\,619 samples.

\noindent
\textbf{Test Datasets:} \underline{IEMOCAP~\cite{busso2008iemocap}:} We follow the most frequently used category selection (angry, happy+excited, neutral, and sad) to build the test dataset. Thus, the dataset contains 5\,531 utterances.
\underline{RAVDESS~\cite{livingstone2018ryerson}:} The speech data consists of 1\,440 utterances with 8 expressions (\textbf{angry, calm, disgust, fearful, happy, neutral, sad}, and \textbf{surprise}).
\underline{CREMA-D~\cite{cao2014crema}:} The dataset consists of 7\,442 clips from 91 actors, covering facial and vocal emotional expressions in sentences with a diverse range of basic emotional states (\textbf{anger, disgust, fear, happy, neutral}, and \textbf{sad}).
\underline{TESS~\cite{pich2020tess}:}  The dataset contains 2\,800 clips featuring seven emotions:\textbf{anger, disgust, fear, happiness, neutral, pleasant surprise}, and \textbf{sadness}. 
\underline{FAU-Aibo~\cite{batliner2008releasing}:} The FAU-Aibo Emotion Corpus is a German speech emotion database containing children speech in the ages 6 to 10\,years. It includes a training set of 9\,959 speech chunks and a test set of 8\,257 chunks. 
The original 11 labels are mapped to
a) five categories (\textbf{angry, emphatic, neutral, positive}, and \textbf{rest}), and
b) two-categories, labelled as \textbf{`non-negative'} and \textbf{`negative'}.~\cite{schuller2009interspeech}.
\underline{ALC~\cite{schiel2012alcohol}:} The Alcohol Language Corpus contains German speech collected under a systematic intoxication test. The INTERSPEECH 2011 Speaker State Challenge~\cite{schuller2011interspeech} selected part of the ALC to obtain a gender and age balanced dataset. 1\,620 data with the label \textbf{`not intoxicated with alcohol'} and 1\,380 labelled as \textbf{`intoxicated with alcohol'} are contained in the test set.
\underline{SLD~\cite{burkhardt2010database}:} The Speaker Likability Database is a subset of the German telephone speech dataset aGender and contains 800 audio chunks each for the labels \textbf{`likable'} and \textbf{`non-likable'}.

\noindent
\textbf{Experimental setup:}
During training, all audio sequences are clipped or padded to achieve a consistent 5-second duration. Both the audio and local feature extractor are frozen while text branches are learnable with a learning rate of $1e-5$. Meanwhile, the projection layers and other parameters employ a learning rate of $1e-3$. We set the mix gamma $\gamma$ to 0.1 and fusion factor $\beta$ to 0.5. ParaCLAP shares the same training recipe for a fair comparison.
All models are trained with an Adam optimizer and a batch size of 32 with training epochs is set to 10. 
During inference, we use the annotated labels provided by the datasets as the text query.

\noindent
\textbf{Baselines:} Our main baseline is ParaCLAP~\cite{Jing24-PTA}, a CLAP-style model trained with identical data and experimental parameters as our proposed SmoothCLAP.
As additional baselines, we utilised the publicly-available CLAP~\cite{Elizalde23-CLA} and Pengi models~\cite{Deshmukh24-PAA} without further fine-tuning or retraining.
Note that both models have been trained on different datasets and feature different underlying components.
CLAP uses CNN14 as its audio encoder and BERT as its text encoder.
Pengi uses HTSAT as its audio encoder and features GPT-2 as its generative decoder; zero-shot inference with Pengi is done by comparing the similarity of the generated caption with that of the target label.

\subsection{Results}
\begin{table}[t]
\centering
\caption{The Unweighted Average Recall (UAR) results on 8 affective computing tasks, comprising both English and German language-based datasets. \textbf{Bold} marks the best performance while \underline{underline} represents the second best.}
\label{tab:emods}
\begin{tabular}{l|c c c c}
\hline
\multirow{2}{*}{\rule{0pt}{4.7ex}Dataset} & \multicolumn{4}{c}{\rule[-1.2ex]{0pt}{3.6ex}Unweighted Average Recall (UAR)} \\ \cline{2-5}

&\thead{CLAP} & \thead{Pengi}&\thead{ParaCLAP}& \thead{\sclap} \\
\hline
\multicolumn{1}{l|}{\rule{0pt}{2.8ex}IEMOCAP (4cl/en)} & .353 &.345 & \underline{.600}& \textbf{.606}\\
\midrule

\multicolumn{1}{l|}{RAVDESS (8cl/en)} &\underline{.199} & .148 &\textbf{.228}&  .175 \\
\midrule

\multicolumn{1}{l|}{CREMA-D (6cl/en)} &.230 & \underline{.245} &.177&  \textbf{.266} \\
\midrule

\multicolumn{1}{l|}{TESS (7cl/en)} &\underline{.232} & .177 &.170&\textbf{.275}\\
\midrule

\multicolumn{1}{l|}{FAU Aibo (2cl/de)} &.500 & .470 &\underline{526}&  \textbf{.555} \\
\midrule

\multicolumn{1}{l|}{FAU Aibo (5cl/de)} &\textbf{.211} & .185 &.197& \underline{.204}  \\
\midrule

\multicolumn{1}{l|}{ALC (2cl/de)} &\underline{.511} & .473 &.537&  \textbf{.541} \\
\midrule

\multicolumn{1}{l|}{SLD (2cl/de)} &.472 & .485 &.\textbf{507}&  \underline{.496} \\

\bottomrule
\end{tabular}
\end{table}

Our main results are presented in \cref{tab:emods}.
\sclap generally outperforms prior CLAP variants, obtaining the best performance on 5 out of 8 tasks, and showing the second-best performance on 2 out of the 3 remaining ones.
Beyond emotional labels conforming to Ekman's ``Big-6'' scheme (IEMOCAP, RAVDESS, CREMA-D, and TESS), SmoothCLAP shows strong zero-shot performance on the more niche labels of FAU Aibo as well as on the two non-emotional tasks of ALC and SLD.
This generalisation demonstrates that the proposed soft-target supervision generalizes beyond emotion recognition, suggesting broader applicability of the method.
Moreover, since training was conducted exclusively on English data, the competitive results on German corpora further highlight SmoothCLAP’s zero-shot cross-lingual transferability.
The fact that SmoothCLAP underperforms on 3/8 tasks can be seen as another demonstration of the ``no free lunch'' theorem -- no one algorithm is a panacea.

\begin{figure}
    \centering
    \includegraphics[trim=0.8cm 2.0cm 3.0cm 1.3cm, clip, width=\linewidth]{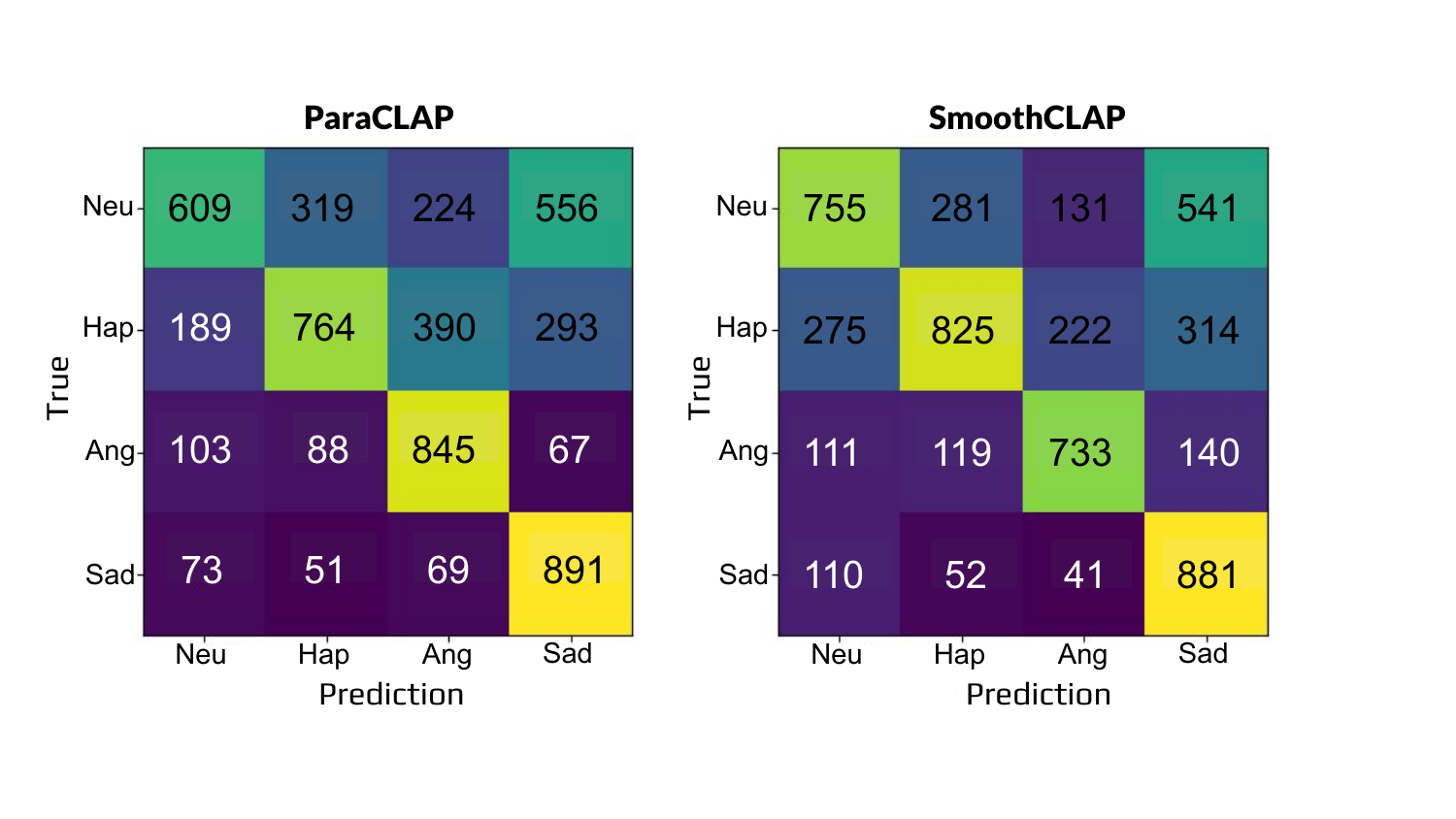}
    \caption{Confusion matrices on IEMOCAP for ParaCLAP (left) and \sclap (right).}
    \label{fig:cm}
\end{figure}

\cref{fig:cm} presents the confusion matrices for IEMOCAP for ParaCLAP and \sclap.
While the two models achieve a similar UAR ($.600$ vs $.606$), they exhibit a very different pattern of misclassifications.
Specifically, \sclap shows a stronger bias towards neutral (reflected in more samples classified as neutral irrespective of their actual label).
We hypothesize that this is a side-effect of the different training strategy which smooths labels.
A stronger preference for neutral reflects a tendency to err on the side of caution; this type of misclassification is more palatable for human users compared to conflating different emotions.

\begin{table}[t]
\centering
\caption{Performance of local feature encoders across test datasets. \textbf{Bold} marks the best performance while \underline{underline} represents the second best.}
\label{tab:lfe}
\resizebox{\columnwidth}{!}{%
\begin{tabular}{lcccc}
\toprule
\textbf{LocFeat Enc.} & \textbf{IEMOCAP} & \textbf{RAVDESS} & \textbf{CREMA-D} & \textbf{TESS} \\
\makecell[l]{Wav2Vec2.0-Emo}~\cite{Wagner23-DOT} & \textbf{.606} & .175 & \textbf{.266} & .275 \\
Wav2Vec2.0-L~\cite{Baevski20-WAF}                   & \underline{.594} & .173 & \underline{.260} & \underline{.368}\\
WavLM-L~\cite{Chen22-WLS}        & .595 & \underline{.212} & .201 & .267 \\
HuBERT-L~\cite{Hsu21-HSS}        & .574 & \textbf{.260} & .259 & \textbf{.433} \\
\bottomrule
\end{tabular}}
\end{table}

\textbf{Local feature extractor: }\cref{tab:lfe} presents the impact of different local feature encoders across the four English emotion datasets. From the results, we see that no single encoder dominates across all datasets; rather, performance varies with the nature of emotional expression. These results show that the choice of local feature extractor significantly influences SmoothCLAP’s ability to capture intra-modal similarity.
They further illustrate how a comparison across multiple datasets is important to gauge the suitability of one feature extractor over others.

\textbf{Mix gamma $\gamma$ and fusion factor $\beta$: }To identify the optimal hyperparameter settings for the mix weight $\gamma$ and the fusion factor $\beta$, we conducted a grid search on the IEMOCAP dataset. Both factors were varied in steps of 0.1 within the range $(0,1)$. 
While $\gamma$ did not show any clearly identified patterns, performance was lower as values of $\beta$ increased.
This indicates that bigger deviations from the identity target (as in ParaCLAP) are detrimental to system performance and that the smoothing factor should be kept to a low magnitude.
The optimal configuration $\gamma=0.5$, $\beta=0.1$ was selected based on empirical performance.

\begin{figure}[t]
\begin{minipage}[b]{.50\linewidth}
  \centering
  \centerline{\includegraphics[trim=0.1cm 0.1cm 0.1cm 0.1cm, clip, width=4.0cm]{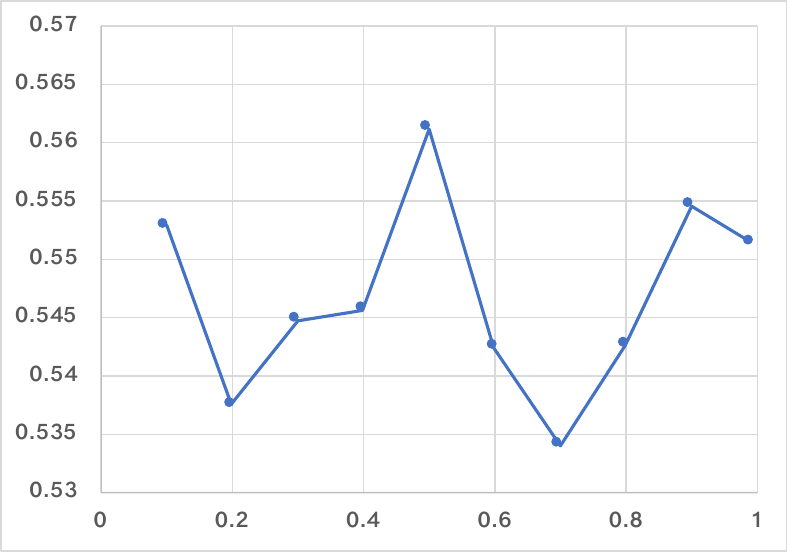}}
  \centerline{(a) Mix gamma $\gamma$}\medskip
\end{minipage}
\hfill
\begin{minipage}[b]{0.50\linewidth}
  \centering
  \centerline{\includegraphics[trim=0.1cm 0.1cm 0.1cm 0.1cm, clip, width=4.0cm]{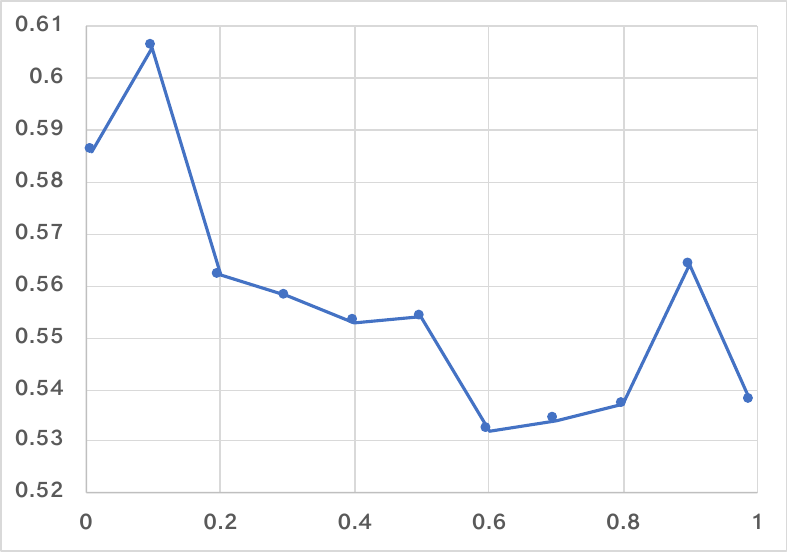}}
  \centerline{(b) Fusion factor $\beta$}\medskip
\end{minipage}
\caption{Performance of mix gamma $\gamma$ and fusion factor $\beta$ }
\label{fig:res}
\end{figure}

\section{Conclusion}
\label{sec:con}

In this work, we proposed \sclap, which incorporates intra-modal similarity derived from computational paralinguistic features, to address the strict one-to-one alignment of paired audio–text samples in conventional CLAP models. Our experiments on both emotional and non-emotional datasets demonstrated the effectiveness of the proposed approach. Nevertheless, the influence of the training data remains an open question, with both quantity and diversity expected to be another key factor.
\section{Acknowledgment}
The authors would like to thank the Munich Center for Machine Learning (MCML), the Munich Data Science Institute (MDSI), and the Reliable AI (relAI) for their support.

\newpage

\bibliographystyle{IEEEbib}
\bibliography{mybib}

\end{document}